\documentclass[12pt,a4paper]{article}
\usepackage{graphicx}
\usepackage[cp1251]{inputenc}
\usepackage[english]{babel}
\usepackage{amssymb, amsfonts, amsmath}
\numberwithin{equation}{section}
\usepackage{amscd}
\date{}
\begin{document}

\title{Solutions of Maxwell equations for admissible electromagnetic fields, in spaces with simply transitive four-parameter groups of motions}

\maketitle

\begin{center}
	\textbf{V.V. Obukhov$^{*}$$^\dag$, S.V. Chervon$^{**}$$^{\dag \dag} $, D.V. Kartashov $^{*}$}
\end{center}

\begin{center}
	\textit{$^{*}$Institute, of Scietific Research and
Development, \\Tomsk State Pedagogical University (TSPU),\\60 Kievskaya St., Tomsk, 634041, Russia; e.mail: obukhov@tspu.edu.ru.}\\[0.2cm]
\end{center}
\begin{center}
	\textit{$^\dag$Laboratory for Theoretical Cosmology, International Center of Gravity and Cosmos, \\Tomsk State University of Control Systems and Radio Electronics (TUSUR),\\36, Lenin Avenue, Tomsk, 634050, Russia}\\[0.2cm]
\end{center}

\begin{center}
	\textit{$^{**}$Laboratory of gravitation, cosmology, astrophysics,\\ Ulyanovsk State Pedagogical University, \\ Lenin's square, 4/5, Ulyanovsk, 432071, Russia,\\ e-mail: chervon.sergey@gmail.com}\\[0.2cm]
\end{center}

\begin{center}
	\textit{$^{\dag\dag}$Physics Department, \\Bauman Moscow State Technical University, \\2-nd Baumanskaya street 5, Moscow, 105005, Russia}\\[0.2cm]
\end{center}

\quad \quad \quad  equation, motion integrals, groups of motions of a space.

            \quad

\abstract {All non-equivalent solutions of vacuum Maxwell  equations are found for the case when space-time manifolds admit simply transitive four-parameter groups of motions $G_4(N)$. The potentials of the admissible electromagnetic fields admit the existence of the algebra of motion integrals of the Hamilton-Jacobi and Klein-Gordon-Fock equations which is isomorphic to the  algebra  of the group operators for the same group $G_4(N)$}.


\section{Introduction}
The subject of our consideration is Maxwell vacuum equations for admissible electromagnetic fields in space-time manifolds $M(G_4)$ (admissible electromagnetic fields are the fields in which the Hamilton-Jacobi and Klein-Gordon-Fock equations admit algebras of motion integrals linear in momentum.
As shown in papers \cite{1}-\cite{2} admissible electromagnetic fields  invariant under action of groups \quad$G_r(N)$. The metrics of the varieties $M(G_4)$, as well as the corresponding  Killing vector fields, has been found in the book by A.Z. Petrov\cite{3}. In the papers \cite{4}-\cite{6} the potentials of admissible electromagnetic fields has been found in coordinate-free form. In the paper \cite{7} all vector tetrad associated with the operators of the groups \quad $G_4(N)$ has been found, as well as all  vector potentials of the  admissible electromagnetic fields. In the papers \cite{9},\cite{10} the potentials of admissible electromagnetic fields has been found for those space-time manifolds, in  which  three-parameter groups of motions act simply transitive on a hypersurface. In the papers \cite{11} a similar problem has been solved for the groups of motions acting  on two-dimensional subspaces of a space-time manifold. Thus, all admissible electromagnetic fields for all space-time manifolds admitting groups of motions $G_r(N) \quad (r\leq 4)$ have been classified.

Currently, there are two methods of exact integration for linear partial differential equations of the first and second orders. Both methods lead to an explicit or implicit procedure for the separation of variables (in the second method, not necessarily spatial ones) in the classical equation of motion of a neutral or charged test particle. For quantum equations these methods are based on the symmetry of the Hamilton-Jacobi equation, which is not surprising, since Killing fields define the integrals of motion namely for this equation.

The first method uses mutually commuting vector and tensor Killing fields, forming a complete set. A large number of articles are devoted to the theory of complete separation of variables (see, for example, articles \cite{12}-\cite{18} and the bibliography presented in them). Spaces in which complete sets exist are called Stackel spaces. Stackel spaces, as well as the theory of complete separation of variables, have found wide application in obtaining and using exact solutions of classical and quantum equations of motion in general relativity and in special theory of relativity (see, for example, \cite{19}-\cite{21}).

The second method (the method of non-commutative integration) can be used in the case when simply transitive four- or three-parameter groups of motions act on $M_4$ or on its hypersurface $V_3$ (see \cite{22}-\cite{27}.

Note that the possibility to exact integrate the free Hamilton-Jacobi equation using one of the methods listed above does not guarantee the successful using of these methods to find exact solutions in the presence of an electromagnetic field for a charged classical test particle, as  for a quantum particle (which obey  Klein-Gordon-Fock,  Dirac-Fock equations and so on). The existence of the required number of integrals of motion that satisfy some additional conditions is a necessary but not sufficient condition for exact or reliable approximate integration of the classical and quantum equations of motion. The classification of fields and metrics in which the specified methods are applicable for such particles is a rather complex problem from a mathematical point of view. Large number of articles are devoted to this problem (see, for example, \cite{28}-\cite{34} and cited in them articles).

Gravitational, electromagnetic and other physical fields, in which the equations of motion of test particles admit the integrals of motion linear and quadratic in momentum, are of significant interest in the theory of gravitation and in relativistic quantum theory, since the study of the features of motion based on the existence of such integrals allows us to obtain important information both about the fields themselves and about the processes occurring in them. There are a large number of articles in which the symmetry of these fields is used to solve various problems in the theory of gravity, in cosmology and in relativistic quantum theory (see, for example,\cite{34}-\cite{58}).

At present, the problem of classifying space-time metrics and admissible electromagnetic fields for groups $G_r(N) r\leq 4$ has been completed. Therefore, we can begin to solve a new classification problem - listing corresponding nonequivalent solutions of the Einstein-Maxwell equations. This problem is preceded by a practically solved classification problem for Maxwell vacuum equations. In the articles \cite{59}-\cite{62} all nonequivalent solutions of these equations for homogeneous spaces and admissible electromagnetic fields have been found. In present paper, this classification is supplemented by solutions of the Maxwell equations for spaces with simply transitive groups of motions $G_4(N)$.

\section{Maxwell equations}

Consider the space-time manifold $M_4$ on which the group of motions $G_4$ acts simply  transitively.
This means that for any two points $ X, X^{'}\simeq M $
in $G_4$ there is a unique shift that transforms $X$ into $X^{'}$. Therefore, there is a smooth one-to-one correspondence between the points of the variety $M_4$ and
elements of the group $G_4$.
The basis $e^A$ in the Lie algebra $\mathfrak{g}$ of the group $G_4$ and its dual
the basis $e_A$ in the coalgebra $\mathfrak{g}^*$ corresponds to non-orthogonal vector tetrads with coordinates $e^\alpha_i$ and $e_\alpha^i$, satisfying the conditions:

\quad

\begin{equation}\label{1}
g_{ik}=e^{\alpha}_ie^{\beta}_k \eta_{\alpha\beta}, \quad g^{ik}=e_{\alpha}^ie_{\beta}^k \eta^{\alpha\beta},\quad e^{\alpha}_ie_{\beta}^i =\eta^{\alpha\gamma} \eta_{\alpha\gamma} =\delta^{\alpha}_\beta, \quad e^{\alpha}_i e^k_{\beta}=\delta^k_i, \quad;
\end{equation}
\begin{equation}\label{2}
e_\alpha^i e^{k}_{\beta,i}-e_{\beta}^i e^{k}_{\alpha,i} = C^\gamma_{\alpha\beta}e^{k}_\gamma,
\end{equation}


\quad

Here $C^\gamma_{\alpha\beta} $ are structure constant of group, \quad $g_{ik}, \quad g^{ik} are $ covariant and contravariant components of the metric tensor on $M_4, \quad\eta_{\alpha\beta} = const$. Let us denote the variables of the holonomic coordinate system by $u^i$. The indices of geometric objects in the holonomic coordinate system $u^i$ will be denoted by the letters $i, j, k = 1, 2, 3, 4$; in nonholonomic (group) reference systems -- $\alpha, \beta, \gamma = 1, 2, 3, 4.$

Admissible electromagnetic fields are those fields, in which the Hamilton-Jacobi and Klein-Gordon-Fock equations admit an algebra of integrals of motion linear in momentum. The components of the admissible electromagnetic field potential $A_i$ satisfy the system of equations (see \cite{1}):

\quad

\begin{equation}\label{3}
\xi^i_\alpha({\mathbb{A}}_{\beta})_{,i}= C^\gamma_{\alpha\beta}{\mathbb{A}}_{\gamma}. \quad
\end{equation}
Here ${{\mathbb{A}}_{\alpha}}=\xi^k_\alpha A_{k}$.
Thus, the admissible electromagnetic field is invariant under the group $G_4$. Tetrad components of vector and tensor fields have the form:
\begin{equation}\label{4}
A_\alpha = e_\alpha^i A_i,  \quad F_{\alpha\beta} = e^i_{\alpha}e^j_{\beta}F_{ij}.
\end{equation}
In the papers \cite{2},\cite{5}, it was shown that
$$
A_\alpha = const.
$$
From \eqref{2} it follows:
\begin{equation}\label{5}
F_{\alpha\beta}=(e^i_{\alpha}e^j_{\beta}e^\gamma_{j,i}-e^i_{\alpha}e^j_{\beta}e^\gamma_{i,j})A_\gamma=
	(e^i_{\alpha|\beta}e^\gamma_i-e^\gamma_ie^i_{\beta|\alpha})A_\gamma=
	C^\gamma_{\alpha\beta}A_\gamma., \quad (X_{|\alpha}=e^i_\alpha \partial_i)
\end{equation}
$$
	\Rightarrow  \quad
F^{i j}=F^{\alpha\beta}e_{\alpha}^ie_{\beta}^j,\quad
	(F^{\alpha\beta}=\eta^{\alpha\alpha_1}\eta^{\beta\beta_1}F_{\alpha_1\beta_1}=
\eta^{\alpha\alpha_1}\eta^{\beta\beta_1}C^\alpha_{\alpha_1\beta_1}A_\alpha).
$$
Using this, we find next form of Maxwell equations:
$$
	(F^{ij})_{;j}e^\gamma_i=(\frac{1}{2}\frac{g_{,j}}{g}e^i_\alpha e^j_\beta +(e^i_\alpha e^j_\beta)_{,i})e^\gamma_i F^{\alpha\beta}=
	(\frac{1}{2}\frac{g_{,j}}{g}e^j_\beta \delta^\gamma_\alpha +e^i_{\alpha|\beta} e^\gamma_i + \delta^\gamma_\alpha e^e_{\beta,e})F^{\alpha\beta}.
$$
As
$$
	\frac{1}{2}(g^{ki}_{|\beta})=
\eta^{\alpha_1\beta_1}\eta_{\alpha_2\beta_2}(e^k_{\alpha_1|\beta}e^i_{\beta_1}e^{\alpha_2}_ke^{\beta_2}_i)=-e^i_{\beta|\alpha}e^\alpha_i-C^\gamma_{\alpha\beta}e^i_\gamma e^\alpha_i=C^\gamma_{\beta\gamma}-e^i_{\beta,i} \quad (g= det|g_{ij}|),
$$
Maxwell vacuum equations reduces to the following system of algebraic equations for \quad $A_\alpha, \quad \eta_{\alpha\beta}$:
\begin{equation}\label{6}
F^{\gamma\beta}C^\alpha_{\beta\alpha}+C^\gamma_{\alpha<\beta}F^{\alpha\beta} = 0,
\end{equation}
Thus, the integration of Maxwell's vacuum equations for admissible electromagnetic fields in space with simply transitive groups of motions $G_4$ is reduced to listing all nonequivalent sets \quad $A_\alpha, \quad \eta_{\alpha\beta}$\quad that define electromagnetic fields and metrics of space-time according to the formulas \eqref{4} and \eqref{1}. In the following we will use the vector fields of the tetrad \eqref{1}, found in the work \cite{1} and consistent with the Killing vector fields given in the book by A.Z. Petrov \cite{3}.
	
\section{Groups $G_4(I)$}

In the book \cite{3} A.Z.Petrov separately considers two metrics of spaces $M_4$ with group $G_4(I)$.
In the first version, the parameter $c \ne 1$ is any real number. The operator $X_4 = \xi^i_4p_i$ has the form:
\begin{equation}\label{1a}
X_4 = (c-1)u^1p_1 + cu^2p_2 +u^3p_3 + p_4.
\end{equation}
In the second option $c=1$. The operator $X_4 = \xi^i_4p_i$ is represented in the form:
\begin{equation}\label{2a}
X_4 = \alpha p_1 + u^2p_2 +u^3p_3 + p_4 \quad (\alpha =const).
\end{equation}
However, it is easy to verify that for $c=1$ the operator \eqref{2a} is obtained from the operator \eqref{1a} by coordinate transformation:
$$u^1 \Rightarrow u^1 + \alpha u^4, $$
Therefore, it is a special case of the operator \eqref{1a} and there is no need to consider this option separately.
The components of the tetrad of vectors $e^A$, $e_A$ can be represented as matrices:

\quad

\begin{equation}\label{3a}
	e_\alpha^i=\begin{pmatrix}
		0 & \exp(c u_4) & 0 & 0\\
		0 &-u_1\exp(u_4) & \exp(u_4) & 0 \\
		\exp(c_1 u_1) & 0 & 0 & 0 \\
		0 & 0 & 0 &-1
	\end{pmatrix},
\end{equation}
$$
	e^\alpha_i=\begin{pmatrix}
		0 & 0 &\exp(-c_1 u_4) & 0\\
		\exp(-c u_4) & 0 & 0 & 0 \\
		u_1\exp(-cu_4) &\exp(-u_4) & 0 & 0 \\
		0 & 0 & 0 & -1
	\end{pmatrix} \quad  (c_1=c-1)
$$ (see. \cite{1}).

Substituting \eqref{4aa} into \eqref{6}, we obtain the system of Maxwell equations in the form:
\begin{equation}\label{4ab}
F^{2 3}=cF^{14}, \quad (1-2c)F^{24}=0,\quad (c+1)F^{34}=0.
\end{equation}
Let us denote:
\begin{equation}\label{4ba}
F^{\alpha 4}=\eta^{44}f^\alpha; \quad \gamma_1 = c \alpha_1, \quad \gamma_2 = \alpha_2, \quad\gamma_3 = c_1 \alpha_3.
\end{equation}
Then from \eqref{4bb} it follows:
\begin{equation}\label{4bb}
f^{\alpha}=\eta^{\alpha\beta}\gamma_\beta \Rightarrow \gamma_\alpha =\eta_{\alpha\beta}f^\beta.
\end{equation}
Let us find the component \quad $F^{23}$:
$$
F^{23}=\eta^{2\alpha}\eta^{3\beta}C^\gamma_{\alpha\beta}\alpha_\gamma =\eta^{2\alpha}\eta^{3\beta}(\delta^\gamma_1(\varepsilon^{23}_{\alpha \beta}+c\varepsilon^{14}_{\alpha \beta})+\delta^\gamma_2 \varepsilon^{24}_{\alpha \beta}+\delta^\gamma_3 c_1 \varepsilon^{34}_{\alpha \beta}) = \frac{\alpha_1\eta_{11}}{\eta}.
$$
Using this, we represent Maxwell equations in the form of the following system of algebraic equations:
\begin{equation}\label{6a}
	\left\{\begin{array}{ll}
		\alpha_1\eta_{1 1} - c\eta^{44}\eta f^1=0, \quad c\alpha_1 =\eta_{1\alpha}f^\alpha \cr
(c+1)f^3 = (2c-1)f^2 =0.		
\\\end{array}\ \right.
\end{equation}
The solution to the system of equations \eqref{6a} depends on the value of the parameter $c$.

\quad

1. Let $c\ne 0,\quad \frac{1}{2},
\quad -1$.
\quad
From \eqref{6a} the solution follows:
\begin{equation}\label{7a}
\eta^{44} = \frac{\eta_{11}^2}{\eta c^2}, \quad \alpha_1 = \frac{f\eta_{1 1}}{c},\quad \alpha_2 = f\eta_{1 2}, \quad \alpha_3 = \frac{f\eta_{1 3}}{c_1}.
\end{equation}

\quad

2. Let $c=0$.  From \eqref{6a} the solution follows:
\begin{equation}\label{8a}
\eta^{11}=0, \quad \alpha_2 = f\eta_{1 2}, \quad \alpha_3 = -f\eta_{1 3}.
\end{equation}
$\alpha_1$ - is an arbitrary parameter.

\quad

3. Let $c=\frac{1}{2} \Rightarrow c_1=-\frac{1}{2}, \quad f^3=0 $. Then from \eqref{6a} it follows:
\begin{equation}\label{8a}
\alpha_1 = 2(\eta_{11}f^1 + \eta_{12}f^2), \quad \alpha_2 = \eta_{12}f^1 + \eta_{22}f^2, \quad \alpha_3 = -2(\eta_{13}f^1 + \eta_{23}f^2).
\end{equation}
$$
2\alpha_1\eta_{11}= \eta\eta^{44}f^1 \Rightarrow f^1(\eta\eta^{44}-4\eta_{11}^2) = 4\eta_{11}\eta_{12}f^2.
$$
Obviously, due to the Lorentzian signature of the metric,
$$
\eta\eta^{44}-4\eta_{11}^2 \ne 0.
$$
Hence the solution can be represented as:
\begin{equation}\label{9a}
\alpha_1 = 2a\eta_{12}\eta^{44}, \quad \alpha_2=a\eta(\eta_{22}\eta^{44}-4\eta_{11}\eta^{33}), \quad \alpha_3=-2a
(\eta_{23}\eta^{44}+4\eta_{11}\eta^{23}),
\end{equation}
$a$ is an arbitrary parameter.

\quad

4. Let $c=-1\Rightarrow f^2=0 $. Then from \eqref{6a} it follows:
\begin{equation}\label{8a}
\alpha_1 = -(\eta_{11}f^1 + \eta_{13}f^3), \quad \alpha_2 = \eta_{12}f^1 + \eta_{23}f^3, \quad \alpha_3 = -\frac{1}{2}(\eta_{13}f^1 + \eta_{33}f^3).
\end{equation}
$$
f^1(\eta_{11}^2-\eta\eta^{44}) = -\eta_{11}\eta_{13}f^3.
$$
From here:
$$
f^1 = \frac{f}{\eta}\eta_{11}\eta_{13}, \quad f^3=\frac{f}{\eta}(\eta\eta^{44}-\eta_{11}^2).
$$
The solution can be represented as:
\begin{equation}\label{9a}
\alpha_1 = 2f\eta_{13}\eta^{44}, \quad \alpha^2=-2f(\eta_{11}\eta^{23}+\eta_{23}\eta^{44}), \quad \alpha^3=f
(\eta_{11}\eta^{22}+\eta_{33}\eta^{44}),
\end{equation}
$a$ - is an arbitrary parameter.

\section{Groups $G_4 (II)$}

The components of the tetrad of vectors $e^A$, $e_A$ can be represented as matrices:
\begin{equation}\label{1b}
	e_i^\alpha=\begin{pmatrix}
		0              & u_4 \exp(-u_4) & -exp(-u_4) & 0\\
		exp(-2 u_4)    &0               &0           & 0 \\
		u_1exp(-2 u_4) & exp(-u_4)      &0           &0 \\
		0              & 0              & 0          &-1
	\end{pmatrix},
\end{equation}
$$
e^i_\alpha=\begin{pmatrix}
	0 		  & exp(2 u_4)        & 0            & 0\\
	0 		  &-u_1\exp(u_4)      &\exp(u_4)     & 0 \\
	-exp(u_4) &-u_1 u_4 \exp(u_4) &u_4 \exp(u_4) & 0 \\
	0         & 0                 & 0            & -1
\end{pmatrix}.
$$
The structure constants of the groups $G_4(I)$ have the form:
\begin{equation}\label{2b}
	C^\gamma_{\alpha \beta}=\delta^\gamma_1(2\varepsilon^{1 4}_{\alpha \beta}-\varepsilon^{2 3}_{\alpha \beta}) + \delta^\gamma_2 \varepsilon^{2 4}_{\alpha \beta}  + \delta^\gamma_3 \varepsilon^{3 4}_{\alpha \beta}. \quad 
\end{equation}
Using the relations \eqref{1b}, we find the components of the vector potential of the permissible electromagnetic field:
\begin{equation}\label{3b}
	A_{1} = (\alpha_2 u_4-\alpha_3)\exp(-u_4), \quad A_{2}=\alpha_1\exp(-2u_4),
\end{equation}
$$
A_{3}= \alpha_1 u_1 \exp(-2 u_4)+ \alpha_2\exp(-u_4), \quad A_{4}=0.\quad
$$
The Maxwell equations have the form:
\begin{equation}\label{4b}
                     \delta^\alpha_1(F^{2 3}+2F^{1 4})+
3\delta^\alpha_2F^{2 4}+3\delta^\alpha_3F^{3 4}=0 \Rightarrow
\end{equation}
\begin{equation}\label{4bb}
F^{2 4}=0, \quad F^{3 4}=0, \quad F^{2 3}+2F^{1 4}=0.
\end{equation}
Let us denote $F^\alpha =F^{\alpha4}\Rightarrow F^\alpha =\eta^{\alpha\beta}\alpha_\beta.$ Then equations \eqref{4bb} take the form:  $$
	F^{23} + F^{14}= \alpha_1(2\eta^{11}\eta^{4 4}+{\eta^{2 3}}^2-\eta^{2 2}\eta^{3 3})=0, \quad F^{2}= F^{3} =0. \quad \alpha_\beta =\eta_{\alpha 1} F^1.
\quad
$$
The system of equations \eqref{4b} takes the form:
\begin{equation}\label{5b}
(2\alpha_1 \delta^\alpha +\alpha_2 \delta^\alpha_2 +\alpha_3 \delta^\alpha_3)=\delta^\alpha _1 f \quad (F^\alpha = \delta^\alpha _1 f.)
\end{equation}
From the equations \eqref{5b} it follows:
$$2\eta \eta_{44}=\eta_{11}^2 \Rightarrow g > 0, $$
which contradicts the Lorentzian signature of the metric.
Consequently, Maxwell's equation in this case leads to a zero electromagnetic field.

\section{Groups $G_4 (III)$}
The components of the tetrad of vectors $e^A$, $e_A$ can be represented as matrices:
\begin{equation}\label{1c}
	e^i_\alpha=\begin{pmatrix}
		0              &   \exp(2q) \sin\alpha    & 0                     & 0 \\
		\exp(q) \cos (p) & -u^1\exp(q) \cos(p-\alpha) & \exp(q) \cos(p-\alpha)  & 0 \\
		\exp(q) \sin(p)  &  u^1\exp(q) \sin(p-\alpha) & -\exp(q) \sin(p-\alpha) & 0 \\
		0              &  0                       & 0                     &-1
	\end{pmatrix},
\end{equation}
$$
e^\alpha_i=\begin{pmatrix}
	0 		 & \frac{\exp(-q)\sin(p-\alpha)}{\sin\alpha} & \frac{\exp(-q)\cos(p-\alpha)}{\sin\alpha} & 0\\
	\exp(-2q) & 0                                        & 0                                        & 0 \\
	0        & \frac{\exp(-q)\sin(p)}{\sin\alpha}        & \frac{\exp(-q)\cos(p)}{\sin\alpha}        & 0 \\
	0        & 0                                        & 0                                        & -1
\end{pmatrix}.
$$
Here  $p=\sin\alpha,\quad q=\cos\alpha \quad (\alpha = const).$
The structure constants specified by the tetrad \eqref{1b} have the form:
\begin{equation}\label{2c}
C^\gamma_{\alpha \beta}= \delta^\gamma_1(\varepsilon^{2 3}_{\alpha \beta}+2\varepsilon^{1 4}_{\alpha \beta}\cos(\alpha))+\delta^\gamma_2(cos(\alpha)\varepsilon^{2 4}_{\alpha \beta}-\sin(\alpha)\varepsilon^{3 4}_{\alpha \beta})+
\delta^\gamma_3 (\sin(\alpha)\varepsilon^{2 4}_{\alpha\beta}+\cos(\alpha)\varepsilon^{3 4}_{\alpha \beta}).
\end{equation}
Using \eqref{1b}, we find the holonomic components of the vector potential of the admissible electromagnetic field:
\begin{equation}\label{2c}
	A_1=(\alpha_2\sin(p-\alpha)+\alpha_3\cos(p-\alpha))\exp(-q), \quad
	A_2=\alpha_1\exp(-2q)
\end{equation}
$$
A_3=(\alpha_2sin(p)+\alpha_3cos(p))exp(-q).
$$
The Maxwell equations have the form:
\begin{equation}\label{3c}
\delta^\alpha_1(F^{2 3}-2F^{1 4}\cos(\alpha)) -
\delta^\alpha_2(3F^{2 4}\cos(\alpha)+F^{3 4}\sin(\alpha)) +
\delta^\alpha_3(F^{2 4}\sin(\alpha)+F^{3 4}\cos(\alpha))
\end{equation}
From here it follows:
\begin{equation}\label{4c}
	F^{2 4}=F^{3 4}=0,\quad
	F^{2 3}-2F^{1 4}\cos(\alpha)=0,\quad
	F^{2 3}=\alpha_1(\eta^{2 2}\eta^{3 3}-{\eta^{2 3}}^2),
\end{equation}
Let us find the components of the tensor \quad $F^{\alpha\beta}$ included in \eqref{4c}:
$$F^{\alpha 4}=F^\alpha \eta^{4 4}a_\alpha,\quad F^{23}= \frac{\eta_{11}\alpha_1}{2-\eta\cos \alpha},$$
where denoted:
 $$a_1 =2\alpha_1\cos \alpha, \quad a_2 =\alpha_2\cos(\alpha)+\alpha_3\sin(\alpha), \quad a_3 = -\alpha_2\cos(\alpha)+\alpha_3\sin(\alpha)$$
From the Maxwell equations \eqref{4c} it follows:
\begin{equation}\label{5c}
	F^\alpha = \delta^{\alpha}_1f,\quad (4\eta\eta^{44}\cos^2\alpha\-{\eta_{11}}^2)f=0.
\end{equation}
Since \quad $f \ne 0,\quad $ it follows: $$\quad 4\eta\eta^{44}\cos^2\alpha\ - {\eta_{11}}^2 =0, \quad $$ which is impossible due to the Lorentzian signature of the metric.

\section{Groups  $G_4 (IV)$}
The components of tetrad vectors can be represented in the form of the matrices:
\begin{equation}\label{1d}
	e^i_\alpha=\begin{pmatrix}
		1 & 0        & 0         & 0\\
		0 & \exp(u_1)& 0         & 0 \\
		0 & 0        & \exp(-u_4)& 0 \\
		0 & 0        & 0         & 1
	\end{pmatrix}, \quad
e^\alpha_i=\begin{pmatrix}
	1& 0        & 0        & 0\\
    0& exp(-u_1)& 0        & 0 \\
    0& 0        & exp(u_4) & 0 \\
    0& 0        & 0        & 1
\end{pmatrix}.
\end{equation}
We write the structure constants in the form:
$$
C^\gamma_{\alpha\beta}=\delta^\gamma_2 \varepsilon^{1 2}_{\alpha\beta}+\delta^\gamma_3 \varepsilon^{3 4}_{\alpha\beta} \quad \Rightarrow \quad
	C^\gamma_{\beta\gamma}=\delta^1_\beta-\delta^4_\beta.
$$
Using \eqref{1d}, we find the holonomic components of the vector potential of the admissible electromagnetic field:
\begin{equation}\label{2c}
	A_1=\alpha_1, \quad
	A_2=\alpha_2\exp(-u^1), \quad A_3=\alpha_3exp(u^4), \quad A_4 =0.
\end{equation}
The Maxwell equations have the form:
\begin{equation}\label{2d}
(\delta^\alpha_1 + \delta^\alpha_4)F^{14}
-\delta^\alpha_2F^{2 4} -\delta^\alpha_3 F^{13}=0.
\end{equation}
From here it followes:
\begin{equation}\label{3d}
	F^{1 4}=F^{2 4}=F^{1 3}=0.
\end{equation}
Let us find the components of the tensor \quad $F^{\alpha\beta}$ included in \eqref{3d}:
$$F^{1 4}=\alpha_3 \eta^{13} \eta^{4 4},\quad F^{24}= \alpha_3 \eta^{23} \eta^{4 4},\quad F^{1 3}=\alpha_2(\eta^{11} \eta^{23}- \eta^{12} \eta^{13}).
$$
Since \quad $\eta^{44} \ne 0,$ \quad and also \quad $|\alpha_2|+|\alpha_3| \ne 0,$, \quad we obtain the following solutions to the equations \eqref{3d}:

\quad

1. \quad $\alpha_3 = \eta_{23} = 0$.

\quad

2. $\eta^{13} =\eta^{23} = 0,$ \quad
$\alpha_2, \quad \alpha_3 $ \quad  are arbitrary parameters.

\section{Groups $G_4 (V)$}

The components of tetrad vectors can be represented in the form of the matrices:

\begin{equation}\label{1f}
	e^\alpha_i=\begin{pmatrix}
		1 & 0                  & 0                   & 0\\
		0 & \cos u_4 \exp(-u_1)& \sin u_4 \exp(-u_1) & 0 \\
		0 & \sin u_4 \exp(-u_1)& -\cos u_4 \exp(-u_1)& 0 \\
		0 & 0                  & 0                   & 1
	\end{pmatrix},
\end{equation}
$$
e^i_\alpha=\begin{pmatrix}
	1& 0                 & 0                  & 0 \\
	0& \cos u_4 \exp u_1 & \sin u_4 \exp u_1  & 0 \\
	0& \sin u_4 \exp u_1 & -\cos u_4 \exp u_1 & 0 \\
	0& 0                 & 0                  & 1
\end{pmatrix}.
$$	
We write the structure constants in the form:
$$
C^\gamma_{\alpha\beta}=-\delta^\gamma_2(\varepsilon^{1 2}_{\alpha\beta}+\varepsilon^{34}_{\alpha\beta})
-\delta^\gamma_3( \varepsilon^{1 3}_{\alpha\beta}-\varepsilon^{24}_{\alpha\beta}) \quad \Rightarrow \quad
	C^\gamma_{\beta\gamma}=2\delta^1_\beta.$$
Using \eqref{1f}, we find the holonomic components of the vector potential of the admissible electromagnetic field:
\begin{equation}\label{2f}
	A_1=\alpha_1,\quad
	A_2=(\alpha_2cos(u_4)+\alpha_3sin(u_4))exp(u_1),\quad
	A_3=(\alpha_2sin(u_4)-\alpha_3 cos(u_4))exp(u_1)
\end{equation}
The Maxwell equations have the form:
\begin{equation}\label{3f}
\delta^\alpha_2(F^{12} - F^{34}) + \delta^\alpha_3(F^{13} + F^{24})
+2\delta^\alpha_4F^{1 4}=0.
\end{equation}
Hence, it follows:
\begin{equation}\label{4f}
	F^{1 4}=0,\quad F^{1 2}=F^{3 4}, \quad F^{13}=-F^{2 4}.
\end{equation}
Let us find the components of the tensor \quad $F^{\alpha\beta}$ included in \eqref{4f}:
\begin{equation}\label{5f}
\eta F^{12} = \alpha_3\eta_{23} -\alpha_2\eta_{33}, \quad \eta F^{13} = \alpha_2\eta_{23} -\alpha_3\eta_{22}.
\end{equation}
$$
F^{14} =\eta^{44}( \alpha_3\eta^{12} -\alpha_2\eta^{13}), \quad   F^{24} =\eta^{44}( \alpha_3\eta^{22} -\alpha_2\eta^{23}), \quad F^{34} =\eta^{44}( \alpha_3\eta^{23} -\alpha_2\eta^{23}). \quad
$$
Without loss of generality, the solution to the first equation of the system of equations \eqref{4f} can be written as:
\begin{equation}\label{6f}
\alpha_3 = a\alpha_2, \quad \eta^{13}=a\eta^{12}.
\end{equation}
Therefore, the Maxwell equations \eqref{4f} can be written as follows:
\begin{equation}\label{7f}
\eta^{11}(\eta^{23}+a\eta^{33})-a(a^2 + 1){\eta^{12}}^2 = \eta^{44}(a\eta^{22}-\eta^{23}),
\end{equation}
$$
\eta^{11}(\eta^{22}+a\eta^{23})-(a^2 + 1){\eta^{12}}^2 =
\eta^{44}(\eta^{33}-a\eta^{23}).
$$
The first equation in this system can be replaced by next linear combination with the second equation:
\begin{equation}\label{8f}
(\eta^{11} +\eta^{44})(a(\eta^{33} - \eta^{22}) - (a^2-1)\eta^{23}) = 0.
\end{equation}
Let's find all nonequivalent solutions of the system \eqref{8f} \eqref{8f}.

\quad

1. $ \eta^{11} = - \eta^{44} \Rightarrow \eta^{33} = -(\eta^{22} + \frac{a^2 + 1}{\eta^{44}}{\eta^{12}}^2).$

\quad

2. $a(\eta^{33} - \eta^{22}) - (a^2-1)\eta^{23} = 0.$
The second equation of the system \eqref{7f} can be represented as:

\begin{equation}\label{8ff}
\eta^{44}(\eta^{23}+a\eta^{22}) = a(-\eta^{11}(a\eta^{23}+\eta^{22}) + (a^2 + 1){\eta^{12}}^2)
\end{equation}

Let us show that
\begin{equation}\label{8ffff}
a(\eta^{22}-\eta^{23}) \ne 0.
\end{equation}

a. If $a=0$ from equation\eqref{8ff} it follows:
$$
\eta^{23}\eta^{44}=0 \Rightarrow \eta^{23} = 0.
$$
From  equation \eqref{8f} we obtain:
$$
\eta^{33}\eta^{44}=\eta^{11}\eta^{22} - {\eta^{12}}^2  \quad \Rightarrow g>0,
$$
which does not correspond to the Lorentzian signature.

b. Let us show that condition $\eta^{23}=a\eta^{22}$ leads to degeneration of the metric, as in this case  $g =0$. Indeed from equation \eqref{8ff} it follows:
$$
\Rightarrow (a^2+1)(\eta^{11}\eta^{22} - {\eta^{12}}^2) =0, \quad \eta^{33 }=a\eta^{22} \Rightarrow \eta =0.
$$

So the single solution remains:
From \eqref{8f} it follows
\begin{equation}\label{9f}
\eta^{33}=\eta^{22} +\frac{a^2 -1}{a}\eta^{23}, \quad
\eta^{44}=\frac{a}{a\eta^{22}-\eta^{23}}(\eta^ {11}(\eta^{22}+a\eta^{23})-(a^2 +1){\eta^{12}}^2)).
\end{equation}

\quad

\section{Unsolvable Groups}

\subsection{Groups $G_4 (VII)$}

The components of tetrad vectors can be represented in the form of matrices:
\begin{equation}\label{1g}
e^i_\alpha=\begin{pmatrix}
1 & 0 & 0 & 0\\
\text{\~{u}}_1 & -u_2 & -1 & 0 \\
\text{\~{u}}^2_1& (1-2\tilde{u}_1u_2)& -2\text{\~{u}}_1 & 0 \\
\tilde{\varepsilon} & 0 & 0 & -1
\end{pmatrix},
\end{equation}
$$
e^\alpha_i=\begin{pmatrix}
1 & 0 & 0 &0\\
\text{\~{u}}^2_1 & -2\text{\~{u}}_1 & 1 &0\\
u_1(1-\text{\~{u}}_1u_2)& (2\text{\~{u}}_1u_2-1)& -u_2 &0\\
\tilde{\varepsilon} & 0 & 0 &-1
\end{pmatrix}.
$$
  \begin{equation}
\tilde{\varepsilon} =-1, 0, +1 \quad \text{\~{u}}_1=u_1+\varepsilon u_4
\end{equation}
This form of the tetrad is consistent with the set of Killing vector fields for this group, given in the book by A.Z. Petrov:
\begin{equation}\label{2g}
X_1=(p_1 -{u^2}^2p_2 - 2u^2p_3)\exp{-u^3}, \quad X_2 = p_3 , \quad X_3p_2 \exp{u^3}, \quad X_4 = p_1 + p_4 \varepsilon.
\end{equation}
However, if you carry out the coordinate transformation:
  \begin{equation}\label{3g}
u^4 \Rightarrow u_4 + \varepsilon u^1,
  \end{equation}
$\varepsilon$ in \eqref{2g} can be set to zero $\Rightarrow \tilde{\varepsilon} = 0 \Rightarrow \eta_{\alpha\beta} = const.$
Obviously, this statement is also true for the group $G_4(VIII).$ We write the structure constants and components of the electromagnetic field vector in the form:
\begin{equation}\label{4g}
	C^\gamma_{\alpha\beta}=\delta^\gamma_1\varepsilon^{1 2}_{\alpha\beta}+2\delta^\gamma_2\varepsilon^{1 3}_{\alpha\beta}+ \delta^\gamma_3\varepsilon^{2 3}_{\alpha\beta}, \quad C^\beta_{\alpha\beta} =0;
\end{equation}
\begin{equation}\label{5g}
	A_1=\alpha_1,\quad A_2=\alpha_1 {u^1}^2-2\alpha_2 u^1+\alpha_3,
\end{equation}
$$
	A_3=\alpha_1u_1(1-{u}^1 u^2)+(2u^1 u^2-1)\alpha_2-\alpha_3u^2,\quad
	A_4=\alpha_4.
$$
The Maxwell equations have the form:
\begin{equation}\label{6g}
		F^{1 2}=F^{1 3}=F^{2 3}=0.
\end{equation}
Let's denote:
$$
a^1 = \alpha_3, \quad a^2 = -2\alpha_2, \quad a^3 = \alpha_1. \quad
$$
Then the system of equations \eqref{6g} can be represented as:
$$
	\eta_{\alpha\beta}a^\beta = 0.
$$
From here it follows:
\begin{equation}\label{7g}
	\eta (\alpha_1^2 + \alpha_2^2 + \alpha_3^2)=0.
\end{equation}
Thus, the system \eqref{6g} has only the trivial solution:
$$A_i = 0.$$

\subsection{Groups $G_4 (VIII)$}

The components of tetrad vectors can be represented in the form of matrices:

\begin{equation}
	|e^i_\alpha|=\begin{pmatrix}
		\cos u^3 & \frac{\sin u^3}{\sin u^1} &
-\frac{\sin u^3 \cos u^1}{\sin u^1}& 0\\
		-\sin u^3 & \frac{\cos u^3}{\sin u^1} & -\frac{\cos u^3\cos u^1}{\sin u^1}&0 \\
		0                   & 0 & 1 & 0 \\
		0                   & 0          & 0               & 1
	\end{pmatrix},
\end{equation}
$$
e^\alpha_i=\begin{pmatrix}
	\cos u^3 & -\sin u^3 & 0 & 0\\
	\sin u^1\sin u^3 &\sin u^1\cos u^3&
	\cos u^1 &0 \\
	0                   & 0 & 1 & 0 \\
	0                   & 0          & 0                & 1
\end{pmatrix}.
$$
This form of the tetrad is consistent with the set of Killing vector fields for this group, given in the book by A.Z. Petrov.The structural constants and components of the electromagnetic field vector have the form:
\begin{equation}\label{8g}
	C^\gamma_{\alpha\beta}=\delta^\gamma_1\varepsilon^{23}_{\alpha\beta}+\delta^\gamma_2\varepsilon^{31}_{\alpha\beta}+ \delta^\gamma_3\varepsilon^{12}_{\alpha\beta}, \quad C^\beta_{\alpha\beta} =0;
\end{equation}
\begin{equation}\label{9g}
	A_1=\alpha_1\cos u^3 - \alpha_2\sin u^3 ,\quad A_2=\sin u^1(\alpha_1 \sin u^3 + \alpha_2\cos u^3) +\alpha_3 \cos u^1,
\end{equation}
$$
	\quad A_3=\alpha_3,\quad A_4=\alpha_4.
$$
The Maxwell equations have the form:
\begin{equation}\label{10g}
		F^{1 2}=F^{1 3}=F^{2 3}=0.
\end{equation}
Let's denote
$$
a^\alpha = \delta^{\alpha\beta}\alpha_\beta.
$$
Then the system of equations \eqref{10g} can be represented as:
$$
	\eta_{\alpha\beta}a^\beta = 0.
$$
\begin{equation}\label{11g}
	\eta (\alpha_1^2 + \alpha_2^2 + \alpha_3^2)=0.
\end{equation}
Thus, the system \eqref{10g} has only the trivial solution:
$$
A_i = 0.
$$

\section{Conclusion}
 All solutions of Maxwell's vacuum equations for admissible electromagnetic fields are obtained under the condition that the motion groups $G_r(N)\quad (r\leq 4).$ act simply transitively on the space-time metrics. The groups $G_4(VI)$ were not considered because, as shown in \cite{1}, for them the permissible electromagnetic fields vanish. Thus, the problem of classifying admissible electromagnetic fields satisfying Maxwell’s vacuum equations for space-time metrics with motion groups $G_r(N)\quad (r\leq 4)$ has been completely solved (see \cite{2}, \cite{ 59}-\cite{62}).

\quad

FUNDING: The work is supported by Russian Science Foundation, project number N 23-21-00275.

CONFLICTS OF INTEREST: The authors declares no conflict of interest.


\begin{thebibliography}{99}

\bibitem{1}
V.V.Obukhov. Algebras of integrals of motion for the Hamilton-Jacobi and Klein-Gordon-Fock equations in spacetime with a
four-parameter movement group in the presence of an external electromagnetic fields.
Journal of Mathematical Physics. (2022), Vol.63, Issue 2, doi.org/10.1063/5.0080703/

\bibitem{2} V.V.Obukhov Maxwell Equations in Homogeneous Spaces with Solvable Groups of Motions.
Symmetry (2022), 14, 2595. doi.org/10.3390/sym14122595)

\bibitem{3}
A. Z. Petrov "Einstein Spaces", {\em Oxford}. (1969).

\bibitem{4}
A.A.Magazev, "Integrating Klein-Gordon-Fock equations in an extremal
electromagnetic field on Lie groups". Theor.and Math.Phys. {\em 173:3}, 1654-1667, (2012).. doi: 10.1007/s11232-012-0139-x, arxiv.org/abs/1406.5698.

\bibitem{5}
 A. A. Magazev, I. V. Shirokov, Yu. A. Yurevich, “Integrable magnetic geodesic flows on Lie groups”, 156:2 (2008), 189–206; Theoret. and Math. Phys.TMF,(2008), 156:2 1127–1141/ doi.org/10.4213/tmf6240

\bibitem{6}
A.A.Magazev, "Constructing a Complete Integral of the Hamilton–Jacobi Equation on Pseudo-Riemannian Spaces with Simply Transitive Groups of Motions". Math. Phys. Anal Geom. 24, {\em 11}, (2021). https://doi.org/10.1007/s11040-021-09385-3

\bibitem{7}
V. V. Obukhov Hamilton-Jacobi and Klein-Gordon-Fock equations for a charged test particle in space-time with simply transitive
four-parameter groups of motions. J. Math. Phys.(2023) 64, 093507; doi: 10.1063/5.0158054


\bibitem{9}
Obukhov V.V. Algebra of symmetry operators for Klein-Gordon-Fock Equation. {\em Symmetry}. {\bf2021}, {\em 13},  727 (15p.). https://doi.org/10.3390/sym13040727.

\bibitem{10}
Obukhov V.V. Algebra of the symmetry operators of the Klein-Gordon-Fock equation for the case when groups of motions $G_3$ act
transitively on null subsurfaces of spacetime. {\em Symmetry}. {\bf2022}, {\em 14}, (346). https://doi.org/10.3390/sym14020346

\bibitem{11}
Obukhov, V.V., Myrzakulov, K.R., Guselnikova, U.A. et al. Algebras of Symmetry Operators of the Klein–Gordon–Fock Equation for Groups Acting Transitively on Two-Dimensional Subspaces of a Space-Time Manifold. Russ Phys J 64, 1320–1327 (2021). https://doi.org/10.1007/s11182-021-02457-5

\bibitem{12} P.Stackel. Uber die intagration der Hamiltonschen differentialechung mittels separation der variablen.  Math. Ann.
(1897), 49, 145-147.

\bibitem{13}
 L.P.Eisenhart. Separable systems of stackel". Ann.Math., (1934), {\em 35}, 284-305.

\bibitem{14}
Levi-Civita T. Sulla Integraziome Della Equazione Di Hamilton-Jacobi Per Separazione Di Variabili. (1904), Math.Ann. {\em 59}, 383-397.

\bibitem{15}
Jarov-Jrovoy M.S. Integration of Hamilton-Jacobi equation by complete separation of variables method. J.Appl. Math. Mech., (1963) 27,  No 6, 173-219. doi.org/10.1016/0021-8928(63)90122-9.

\bibitem{16}
 V.N. Shapovalov. Symmetry and separation of variables in the Hamilton-Jacobi equation. Sov. Phys.J.(1978), {\em 21}, 1124-1132,.
 doi: 10.1007/BF00894560.

\bibitem{17}
 V.N. Shapovalov. Stackel`s spaces. Sib. Math. J. {\em 20}, 1117-1130, (1979) doi: org/10.1007/BF00971844.

\bibitem{18} W. Miller. Symmetry And Separation Of Variables. Cambridge University Press:Cambridge, (1984), 318 p.

\bibitem{19}
K.Schwarzschild "Uber das Gravitationsfeld eines Massenpunktes nach der Einsteinschen Theorie". Sitzungsberichte der Kouml; niglich Preussischen Akademie der Wissenschaften. {\em 1}, 189-196, (1916).https://ui.adsabs.harvard.edu/abs/1916SPAW.......189S

\bibitem{20}
Roy P. Kerr "Gravitational Field of a Spinning Mass as an Example of Algebraically Special Metrics". Phys. Rev. Lett. {\bf 11}, {\em 5}, 237-238, (1963).. doi/10.1103/PhysRevLett.11.237

\bibitem{21}
A. A. Friedmann. "Uber die Moglichkeit einer Welt mit konstanter negative Krummung des Raumes". Zs. Phys. {\bf 21}, 326-332, (1924). doi.org/10.1007/BF01328280

\bibitem{22}
A.V. Shapovalov, I.V. Shirokov "Noncommutative integration method for linear partial differential equations. functional algebras and dimensional reduction". Theoret. And Math. Phys., {\em 106:1}, 1-10, (1996). doi.org/10.4213/tmf1093.

\bibitem{23}
 A.I.Breev, A.V. Shapovalov "Non-commutative integration of the Dirac equation in homogeneous spaces". Symmetry. {\em 12}, 1867,(2020). doi.org/10.3390/sym12111867


\bibitem{24} Shapovalov, A.V. On Equivalence between Kinetic Equations and Geodesic Equations in Spaces with Affine Connection. {\it Symmetry } {\bf 2023}, 15(4),  905.

\bibitem{25} Shapovalov, A.V.,  Breev, A.I.  Harmonic Oscillator Coherent States from the Standpoint of Orbit Theory
    {\it Symmetry } {\bf 2023}, 15(2), 282.


\bibitem{26} Breev, A.I.; Shapovalov, A.V. Non-commutative integration of the Dirac equation in homogeneous spaces. {\it Symmetry } {\bf 2020}, 12, 1867.

\bibitem{27}
 Breev, A.I.; Shapovalov, A.V. Yang--Mills gauge fields conserving the symmetry algebra of the Dirac equation in a homogeneous space. {\em J. Phys.: Conf. Ser.} {\bf 2014}, 563, 012004.

\bibitem{28}
 B.Carter "New family of Einstein spaces". Phys.Lett.
{\em A.25}, {\em No 9}, 399-400, (1968). doi.org/10.1016/0375-9601(68)90240-5

\bibitem{29} E.Newman, L Tamburino, T Unti. "Empty-Space Generalization of the Schwarzschild Metric". J. Math. Phys. {\bf 4} {\em 5}, 915–923, (1963). doi.org/10.1063/1.1704018

\bibitem{30} V. G. Bagrov and  A. V. Shapovalov and  A. A.Yevsyevich "Separation of variables in the Dirac equation in Stackel spaces".
Classical and Quantum Gravity. {\bf7}, {\em 4}, 517, (1990). doi.org/10.1088/0264-9381/7/4/004

\bibitem{31} V. G. Bagrov and  A. A. Yevsyevich and A. V. Shapovalov "Separation of variables in the Dirac equation in Stackel spaces.2: External gauge fields".  Classical and Quantum Gravity. {\bf8}, {\em 4}, 163-173, (1991). doi.org/10.1088/0264-9381/8/1/016

\bibitem{32}
V.V.Obukhov "Hamilton-Jacobi equation for a charged test particle in the Stackel space of type (2.0)". Symmetry. {\bf12}, (2020), 12891291. doi: 10.3390/sym12081289.

\bibitem{33}
V.V.Obukhov. "Hamilton-Jacobi equation for a charged test particle in the  Stackel space of type (2.1)". Int. J. Geom. Meth. Mod. Phys. {\bf 17}, {\em 14}, {2050186}, (2020). doi: 10.1142/S0219887820501868.

\bibitem{34} V.V.Obukhov "Separation of variables in Hamilton-Jacobi and Klein-Gordon-Fock equations for a charged test particle in the stackel spaces of type (1.1)". Int. J. Geom. Meth. Mod. Phys. {\bf18}, {\em 03}, (2150036), (2021). doi:10.1142/S0219887821500365

\bibitem{35}
A. Mitsopoulos, M.Tsamparlis, G.A. Leon, A. Paliathanasis. New conservation laws and exact cosmological solutions in Brans-Dicke cosmology with an extra scalar field. Symmetry. 13(8):1364,(2021). doi: 10.3390/sym13081364

\bibitem{36}
S.~Capozziello, M.~De Laurentis and S.~D.~Odintsov,
Eur. Phys. J. C \textbf{72} (2012), 2068
doi:10.1140/epjc/s10052-012-2068-0
[arXiv:1206.4842 [gr-qc]].

\bibitem{37}
J.~de Haro, S.~Nojiri, S.~D.~Odintsov, V.~K.~Oikonomou and S.~Pan,
`Finite-time cosmological singularities and the possible fate of the Universe,''
Phys. Rept. \textbf{1034} (2023), 1-114
doi:10.1016/j.physrep.2023.09.003
[arXiv:2309.07465 [gr-qc]].

\bibitem{38}
S.~Nojiri and S.~D.~Odintsov,
``Unified cosmic history in modified gravity: from F(R) theory to Lorentz non-invariant models,''
Phys. Rept. \textbf{505} (2011), 59-144
doi:10.1016/j.physrep.2011.04.001
[arXiv:1011.0544 [gr-qc]].

\bibitem{39}
Camci Ugur, "Integration of the geodesic equations via Noether symmetries", Int. J. Mod. Phys. D. {\bf 31}, {\em 11}, 2240011, (2022).   doi.org/10.1142/S0218271822400119,

\bibitem{40}
 V. Epp, O. Pervukhina "The Stormer problem for an aligned rotator". MNRAS. {\em 474}, 5330-5339, (2018). doi.org/10.1093/mnras/stx3102.

\bibitem{41}
V. Epp, M.A. Masterova. "Effective potential energy for relativistic particles in the field of inclined rotating magnetized sphere". Astrophys M Space Sci. {\em 353}, 473-483, (2014) . doi.org/10.1007/s10509-014-2066-9.

\bibitem{42} Epp, V., Masterova, M.A. Effective potential energy in Stormer’s problem for an inclined rotating magnetic dipole. Astrophys Space Sci 345, 315–324 (2013). https://doi.org/10.1007/s10509-013-1415-4

\bibitem{43}
Y. Kumaran,A. Ovgun "Deflection angle and shadow of the
reissner-nordstrom black hole with higher-order magnetic correction in
einstein-nonlinear-maxwell fields". Symmetry. {\em14}, 2054, (2022). doi.org/10.3390/sym14102054.

\bibitem{44} Cebeciouglu Oktay and Kibarouglu Salih. "Maxwell-modified metric affine gravity". Eur. Phys. J. C. {\bf 81}, {\em 10}, 900.(2021). doi. org/10.1140/epjc/s10052-021-09685-6.

 \bibitem{45}A. Beesham, S.V. Chervon, S. D. Maharaj, A.S. Kubasov. An emergent universe with dark sector fields in a chiral cosmological model. Quantum Matter, v.2, 388-395, 2013.

\bibitem{46}
I.V. Fomin, S.V. Chervon. Reconstruction of GR cosmological solutions in modi?ed gravity theories. Phys.Rev. D100, no.2, 023511, 2019.

\bibitem{47}
I.V. Fomin, S.V. Chervon. Exact and approximate solutions in the Friedmann Cosmology. Russian Phys. J., NY, v.60, issue 30, pp. 427-440, 2017.

\bibitem{48}
 S. Chervon. Inflationary cosmology without restrictions on the scalar field potential.
General Relativity and Gravitation 36:1547-1553, 2004.

\bibitem{49}
S.V. Chervon, M. Novello and R. Triay. Exact cosmology and specification of an inflationary scenario. Gravitation and Cosmology, v.11, No.4, p. 329-344, 2005.

\bibitem{50} Kibaroglu Salih,   "Generalized cosmological constant from gauging Maxwell-conformal algebra". Phys.Lett.B. {\em 10}, 803 135295, (2020) doi.org/10.1016/j.physletb.2020.135295.

\bibitem{51}
Camci Ugur, "Integration of the geodesic equations via Noether symmetries". Int. J. Mod. Phys. D. {\bf 31}, {\em 11}, 2240011,2022)..
doi.org/10.1142/S0218271822400119,

\bibitem{52}
Claudio Dappiaggi, Benito A. Juarez-Aubry, Alessio Marta Ground. "State for the Klein-Gordon field in anti-de Sitter spacetime with dynamical Wentzell boundary conditions". Phys. Rev. D. {\em 105}, 105017, (2022). doi.org/10.1103/PhysRevD.105.105017.

\bibitem{53}
{E.K. Osetrin, K.E. Osetrin and A.E. Filippov},
{Spatially Homogeneous Models Stackel Spaces of Type (2.1)}.
{\it Russian Physics Journal}.(2020), {\bf 63}, N~3, {410-419}.

\bibitem{54}
{E. Osetrin and K. Osetrin},
{Pure radiation in space-time models that admit integration of the eikonal equation by the separation of variables method}.
{\it J. Math. Phys.}  {\bf 58} (2017), N~11, 112504.

\bibitem{55}
 K. Osetrin, E Osetrin. "Shapovalov wave-like spacetimes".
Symmetry. {\em 12}, 1372, (2020).
doi.org/10.3390/SYM12081372..

\bibitem{56}
{K.E. Osetrin, E.K. Osetrin and E.I. Osetrina},
{Gravitational wave of the {Bianchi} {VII} universe: particle trajectories, geodesic deviation and tidal accelerations}.
{\it European Physical Journal C}. {\bf 82} {2022}, N~10, 894.

\bibitem{57}
{K.E. Osetrin, E.K. Osetrin and E.I. Osetrina},
{Geodesic deviation and tidal acceleration in the gravitational wave of the {Bianchi} type {IV} universe}.
{\it European Physical Journal Plus}. {\bf 137} 2022, N~7, 856.

\bibitem{58}
{K. Osetrin, A. Filippov and E. Osetrin},
{The spacetime models with dust matter that admit separation of variables in {Hamilton--Jacobi} equations of a test particle}.
{\it Mod. Phys. Lett.}  {\bf A31} (2016), N~6, 1650027.

\bibitem{59}
V.V.Obukhov. "Maxwell Equations in Homogeneous Spaces for Admissible Electromagnetic Fields". Universe. {\em 8}, (245), (2022). https://doi.org/10.3390/universe8040245

\bibitem{60}
V.V.Obukhov. "Maxwell Equations in Homogeneous Spaces with Solvable Groups of Motions". Symmetry. {\em 14}, 2595, (2022).doi.org/10.3390/sym14122595

\bibitem{61}
 V. V. Obukhov. "Exact Solutions of Maxwell Equations in Homogeneous Spaces with the Group of Motions G3(IX)". Axioms {\em 12}, {\em 135}, (2023). doi.org/10.3390/axioms12020135

\bibitem{62}
V.V.Obukhov "Exact Solutions of Maxwell Equations in
Homogeneous Spaces with the
Group of Motions G3(VIII)". Symmetry. {\em 15}, 648, (2023). doi.org/10.3390/sym15030648

\end{thebibliography}
\end{document}